\documentclass[%
superscriptaddress,
nofootinbib,
amsmath,amssymb,
aps, physrev,
twocolumn, showkeys 
]{revtex4-2}

\usepackage{comment}
\usepackage{graphicx}
\usepackage{dcolumn}
\usepackage{bm}

\usepackage{kotex}
\usepackage{enumitem,graphicx,amsmath,amssymb,color,multirow}
\usepackage[table]{xcolor}
\definecolor{dkgreen}{rgb}{0,0.6,0}
\definecolor{blue-violet}{rgb}{0.54, 0.17, 0.89}

\usepackage{hyperref}
\usepackage{txfonts}

\begin{document}

\preprint{APS/123-QED}


\title{Ecological networks of viable species with degree-dependent interaction}
\author{Hae Seong Lee}
\affiliation{%
Department of Physics, Inha University, Incheon 22212, Republic of Korea
}%
\affiliation{Institute of Quantum Science, Inha University, Incheon 22212, Republic of Korea}

\author{Deok-Sun Lee}
\email{Contact author: deoksunlee@kias.re.kr}
\affiliation{%
School of Computational Sciences, Korea Institute for Advanced Study, Seoul, 02455, Korea
}%
\affiliation{%
Center for AI and Natural Sciences, Korea Institute for Advanced Study, Seoul, 02455, Korea
}%

\author{Sang Hoon Lee}%
\email{Contact author: lshlj82@gnu.ac.kr}
\affiliation{%
Department of Physics, Gyeongsang National University, Jinju, 52828, Korea
}%
\affiliation{%
Research Institute of Natural Science, Gyeongsang National University, Jinju, 52828, Korea
}%

\author{Samir Suweis}
\email{Contact author: samir.suweis@unipd.it}
\affiliation{
Dipartimento di Fisica ``G. Galilei'', Universit{\`a} di Padova, Padua, Italy
}%
\affiliation{
Padova Neuroscience Center, University of Padova, Padua, Italy
}
\affiliation{INFN, Sezione di Padova, Via Francesco Marzolo 8, I-35131 Padova, Italy}%

\author{Hye Jin Park}
\email{Contact author: hyejin.park@inha.ac.kr}
\affiliation{%
Department of Physics, Inha University, Incheon 22212, Republic of Korea
}%
\affiliation{Institute of Quantum Science, Inha University, Incheon 22212, Republic of Korea}

\date{\today}

\begin{abstract}
The generalized Lotka--Volterra (GLV) framework, recently advanced by dynamical mean-field theory, enables the systematic analysis of large ecological networks. 
When combined with structured interaction topologies, previous studies have shown that the 
viability of species, defined by having a positive stationary abundance,  depends on the number of their interacting neighbors or the ``degree.'' While such model studies usually assume that interaction strengths follow an identical distribution across all connected pairs, real ecological communities often exhibit correlations between interaction strength and a species' degree. To capture this overlooked feature, we introduce degree-dependent interaction strengths into a generalized random LV model. We identify two distinct regimes: a hub-favored phase, where highly connected species survive preferentially, and a hub-suppressed phase, where they face higher extinction risks. We analytically derive the phase boundary where these degree-dependent strengths precisely balance the connectivity effect, leaving all species equally susceptible. Crucially, these phases induce opposing shifts in the degree-degree correlation or ``assortativity'' of the 
network of viable species: the hub-favored phase enhances disassortativity by selectively removing the interactions between low-degree species, whereas the hub-suppressed phase reduces it as the interactions 
involving hub species tend to disappear. Ultimately, our findings demonstrate that degree-dependent interactions are a fundamental mechanism not only for shaping species survival, but for naturally reproducing 
the wide range of assortativity values observed in real ecological networks.
\end{abstract}

\maketitle

\section{Introduction}
\label{sec:introduction}
Interaction among species affects the population dynamics, and consequently, the community composition of ecological systems~\cite{may1972will,allesina2012stability,allesina2015stability}. The generalized random Lotka--Volterra (GRLV) model has served as a tool for investigating the role of interaction in shaping ecological communities by considering random interaction strengths~\cite{galla2018dynamically,bunin2017ecological,roy2019numerical,pasqualini2026linking}, and it has been extended to consider more complex interaction structures~\cite{park2024incorporating,poley2025interaction,aguirre2024heterogeneous,azaele2024generalized}.

Despite this extension, previous studies~\cite{galla2018dynamically,bunin2017ecological,roy2019numerical,park2024incorporating,poley2025interaction,aguirre2024heterogeneous} have largely assumed that interaction strengths stem from an identical distribution across the whole system for simplicity. This assumption, however, naturally contradicts empirical observations that the mean interaction strength tends to become smaller for species with more interaction partners or larger \emph{degree} values~\cite{o2010interaction,koch2026many}. It has long been suggested that species with larger degrees tend to have weaker interaction strengths in ecological systems ~\cite{macarthur1955fluctuations,holling1959components,suweis2014disentangling,wootton2016many,gellner2023stable}. This inverse relationship may arise from finite interaction capacities---where limited time, energy, or spatial resources must be partitioned across multiple partners---as well as metabolic constraints in size-structured communities, where highly connected, larger species exert weaker per-biomass effects~\cite{o2010interaction}. Furthermore, network stability necessitates that highly connected ``hubs'' maintain weaker average interactions to dampen population fluctuations and prevent instability~\cite{koch2026many}. 

While mature, observable ecosystems predominantly reside in this negatively correlated regime, investigating the positively correlated case is also essential for understanding the boundaries of ecological viability. Examining scenarios where hubs exert disproportionately strong interactions allows us to model transient perturbations, such as biological invasions, and provides a mechanistic explanation for why such structures are dynamically pruned from stable networks. 
This crucial fact in real ecosystems naturally motivates us to implement \emph{degree-dependent} interaction strengths in the GRLV model and study how this correlation between the degree and the interaction strength affects species survival and the properties of the resulting ecological networks.

To operationalize this concept, we allow species to interact on different types of model networks, where the interaction strengths between species are drawn from Gaussian distributions whose mean and variance are explicitly determined by the degrees of the interacting species. As a result of this formulation, we identify distinct hub-favored and hub-suppressed phases that exhibit distinct behavior in species survival probability. In the hub-favored (hub-suppressed) phase, hub species show higher (lower) survival probability than low-degree species. This phase-dependent difference in survival probability drives opposing shifts in the degree assortativity of the 
resulting network.

This paper is structured as follows. In Sec.~\ref{sec:model}, we introduce the GRLV model with degree-dependent interaction strength. Then, we demonstrate that the survival probability varies with degree for each species in Sec.~\ref{sec:survivalProbability}. In Sec.~\ref{sec:twoPhases}, we identify the hub-favored and hub-suppressed phases with the critical phase boundary, which occurs when the incoming interaction strength scales inversely with a receiving species' degree and degree-dependent effects perfectly balance network connectivity in turn. Then we present the effect of degree-dependent survival probability of species on the degree assortativity of the resulting networks in Sec.~\ref{sec:assortativity}. Finally, we summarize our findings and discuss possible future work in Sec.~\ref{sec:summary}.

\section{Model}
\label{sec:model}
\begin{figure}[t]
    \centering
    \includegraphics[width=0.9\linewidth]{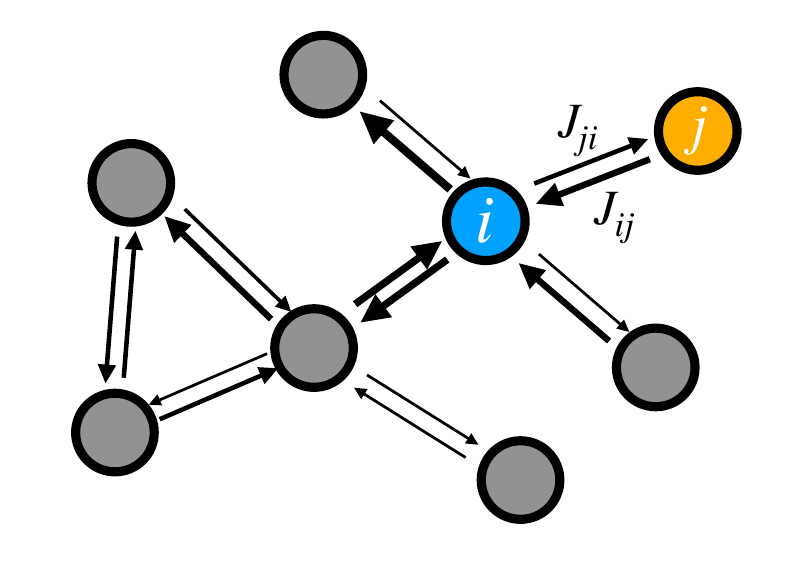}
    \caption{Schematic illustration of the model in Eq.~\eqref{eq:model}. Species are represented as nodes in a network, with directed interactions denoted by arrows. The interaction strength $J_{ij}$, representing the effect of species $j$ on species $i$, is a quenched random variable 
    that may depend on the degrees $k_i$ and $k_j$ of species $i$ and $j$, as defined in Eq.~\eqref{eq:quenchedInteractionStrength}. 
    Interactions are bidirectional ($A_{ij}=A_{ji}$) but their strengths are generally asymmetric ($J_{ij}\neq J_{ji}$).
    }
    \label{fig:schematic}
\end{figure}

We incorporate degree-dependent interaction strengths in the conventional GRLV model~\cite{park2024incorporating,poley2025interaction,aguirre2024heterogeneous} to investigate its impact on the survivability of species. In our model, initially, $S$ species interact via a network with asymmetric bidirectional interactions, as depicted in Fig.~\ref{fig:schematic}. The dynamics of the abundance of species $i$, denoted by $x_i(t)$, is described by the following standard GRLV form:
\begin{equation}
\label{eq:model}
\dot{x}_i(t)=x_i(t)\left[\lambda-x_i(t)-\sum_{j\neq i}J_{ij}A_{ij}x_j(t) \right] \,,
\end{equation}
where $\lambda$ is the growth rate of species $i$, and $A_{ij}$ is an element of the adjacency matrix of the network. The element $A_{ij}$ equals 1 when species $i$ and $j$ are interacting with each other and 0 otherwise. The interaction strength exerted by species $j$ on species $i$ is denoted as $J_{ij}$. We assume that interactions between species are bidirectional, so the adjacency matrix is symmetric ($A_{ij}=A_{ji}$) but interaction strengths are not necessarily symmetric ($J_{ij}\neq J_{ji}$).  The degree $k_i \equiv \sum_j A_{ij}$ of species $i$ represents the number of interacting species. 

The interaction strength $J_{ij}$ is a quenched random variable depending on the degrees of species $i$ and $j$ defined by
\begin{equation}
\label{eq:quenchedInteractionStrength}
J_{ij}=\mu_{ij}(k_i,k_j)+\sigma_{ij}(k_i,k_j)z_{ij} \,,
\end{equation}
where $\mu_{ij}(k_i,k_j)$ and $\sigma_{ij}(k_i,k_j)$ are the mean and standard deviation of $J_{ij}$ given as functions of degrees $k_i$ and $k_j$ of species $i$ and $j$, and $z_{ij}$ is an independent and identically distributed random variable that follows the normal distribution $\mathcal{N}(0,1)$. To capture the topological correlations observed in real ecosystems with minimal mathematical complexity, we extend conventional GRLV models by assuming a power-law scaling for the interaction statistics, depending on the degrees of the interacting species. We define the mean $\mu_{ij}(k_i,k_j)$ and standard deviation $\sigma_{ij}(k_i,k_j)$ as follows:
\begin{equation}
\label{eq:interaction_mean}
\mu_{ij}(k_i,k_j)=\frac{\mu}{K}\left(\frac{k_i}{K}\right)^\alpha\left(\frac{k_j}{K}\right)^\beta \,,
\end{equation}
and
\begin{equation}
\label{eq:interaction_variance}
\sigma^2_{ij}(k_i,k_j)=\frac{\sigma^2}{K}\left(\frac{k_i}{K}\right)^\alpha\left(\frac{k_j}{K}\right)^\beta \,,
\end{equation}
where the baseline mean value $\mu$ determines whether the species interaction is competitive ($\mu>0$) or cooperative ($\mu<0$). On average, the baseline standard deviation $\sigma$ provides the overall variance of interaction strengths, and $K=\sum_{i=1}^S k_i/S$ is the mean degree of species. This specific algebraic choice provides the simplest mathematical form to systematically explore scaling behaviors across the network without introducing additional species-specific parameters. In this study we focus on the competitive environment ($\mu>0$).

The exponents $\alpha$ and $\beta$ in Eq.~\eqref{eq:interaction_mean} and Eq.~\eqref{eq:interaction_variance} govern the correlation between interaction strengths and degrees of interacting species. For an interaction effect of species $j$ on species $i$, $J_{ij}$, both degrees $k_i$ and $k_j$ are involved in determining the strength. To explicitly distinguish the directionality of these topological effects, we utilize both parameters independently for a given interaction $J_{ij}$‬: the exponent $\alpha$ is assigned to the degree of the species that receives the influence, while the exponent $\beta$ is assigned to the degree of the species that exerts the impact.

Thus, from the perspective of a focal species $i$ with degree $k_i$‬, the total incoming influence defined by 
\begin{equation}
    \label{eq:incoming}
    I_i^{\leftarrow}\equiv\sum_{j\in nn(i)} J_{ij}x_j, 
\end{equation}
depends on the individual interaction strengths ($J_{ij}$), the number of interacting species ($k_i$), and their abundances ($x_j$), with $nn(i)$ denoting the set of interacting neighbors. Given that the species receives $k_i$‬ incoming influences and each explicitly scales as $(k_i/K)^{\alpha}$‬,
the total incoming influence is expected to scale as $I_i^{\leftarrow} \propto k_i(k_i/K)^{\alpha} \propto k_i^{\alpha+1}$. Consequently, for $\alpha > -1$‬‬, $I_i^{\leftarrow}$‬ is an increasing function of $k_i$‬, suggesting that hub species experience a higher total incoming interaction strength than non-hub species. On the other hand, when $\alpha < -1$‬‬, $I_i^{\leftarrow}$‬ is a decreasing function of $k$‬, meaning non-hub species experience stronger total incoming interactions than hub species. Exactly at $\alpha = -1$‬‬, this degree-dependent weakening perfectly cancels the topological accumulation of interactions, rendering the total incoming impact independent of degree---a delicate balance that, as we will show later, marks the critical phase boundary for the entire system. 

Similarly, the exponent $\beta$‬ shapes the total outgoing influence,  $I_j^{\rightarrow}\equiv \sum_{i\in nn(j)} x_i J_{ij}$, which is expected to scale as $I_j^{\rightarrow}\propto k_j(k_j/K)^{\beta} \propto k_j^{\beta+1}$‬: $I_j^{\rightarrow}$‬ is an increasing function of $k_j$‬ when $\beta > -1$‬, and a decreasing function when $\beta < -1$‬‬. In the special case of $(\alpha,\beta)=(0,0)$‬‬, the model is equivalent to the previous model~\cite{park2024incorporating} where the interaction strengths between two species are independent of their degrees and are drawn from the same Gaussian distribution. Because the scaling exponents evaluate to $\alpha+1 = 1 > 0$‬ and $\beta+1 = 1 > 0$‬ in this limit, the total interaction strengths scale linearly with degree ($I^{\leftarrow}_i \propto k_i$‬ and $I^{\rightarrow}_j \propto k_j$‬). Consequently, the conventional baseline inherently corresponds to the increasing function case.

\begin{figure*}[t]
    \centering
    \includegraphics[width=.9\linewidth]{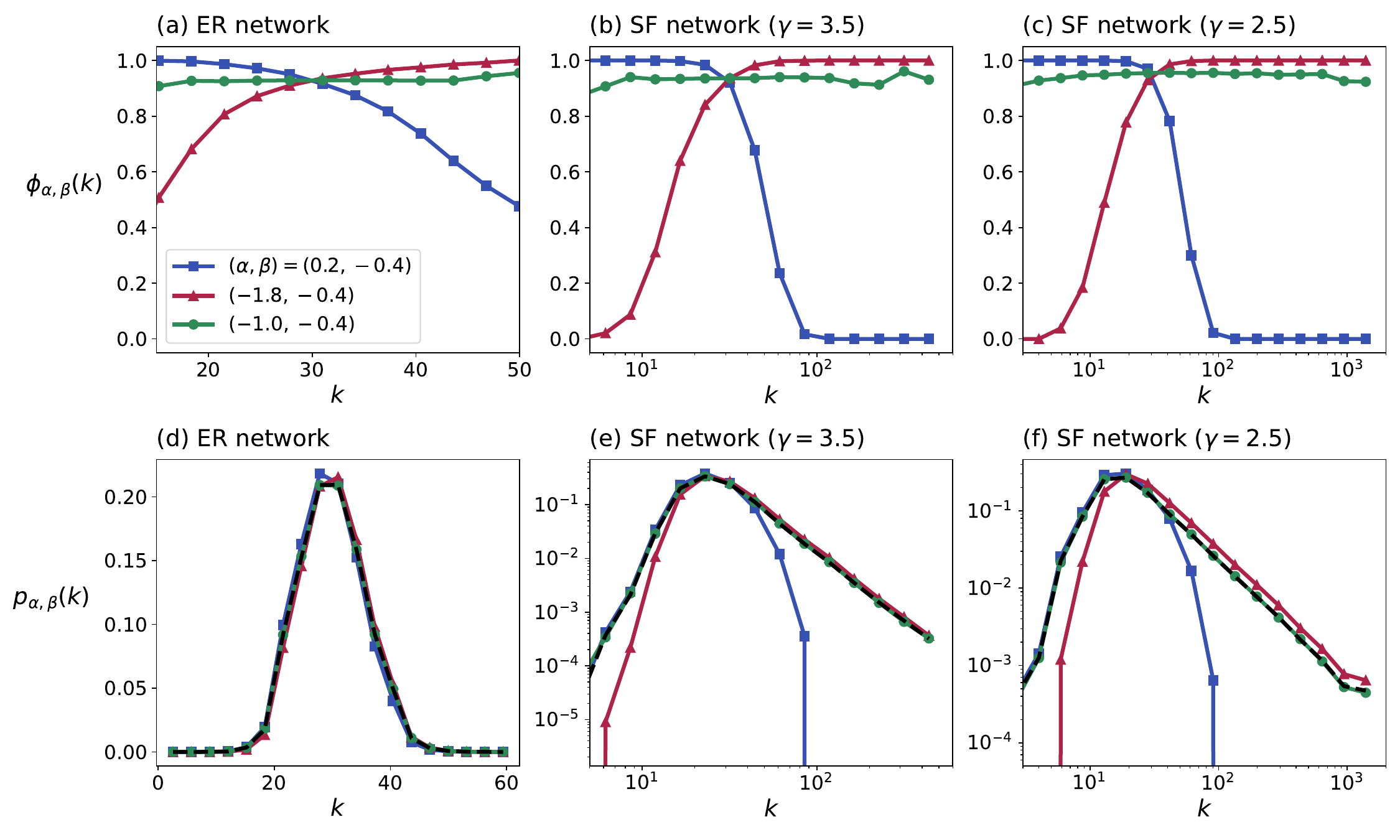}
    \caption{Numerical simulation results of the GRLV model in Eq.~\eqref{eq:model} on ER and static SF networks with $\gamma=3.5$ and $\gamma=2.5$. Panels (a)--(c) show the degree-specific survival probability $\phi_{\alpha, \beta}(k)$, and  panels (d)--(f) show the distribution of the initial degree $p_{\alpha,\beta}^\prime(k)$ of surviving species for 
    $(\alpha,\beta)=(-1.8,-0.4)$, $(-1.0,-0.4)$, and $(0.2,-0.4)$. 
    The survival probability 
    exhibits three distinct behaviors across the given parameter sets for all networks. For $(\alpha,\beta)=(-1.8,-0.4)$ (red triangle), $\phi_{\alpha, \beta}(k)$ increases with $k$, indicating that hub species are more likely to survive than low-degree species. The opposite behavior is observed for $(\alpha,\beta)=(0.2,-0.4)$ (blue square), where $\phi_{\alpha, \beta}(k)$ decreases with $k$, favoring the survival of low-degree species. An unbiased case is observed for $(\alpha,\beta)=(-1.0,-0.4)$ (green circle), where $\phi_{\alpha,\beta}(k)$ is approximately independent of $k$. Accordingly, $p_{\alpha,\beta}^\prime(k)$ 
    nearly coincides with the original degree distribution. The black dashed lines with dots in panels (d)--(f) represent the original degree distribution. 
    }
    \label{fig:survivalProbability}
\end{figure*}

Our mean-field approximation detailed in Appendix~\ref{app:mean-field} gives the stationary abundance of species with degree $k$ as
\begin{equation}
\label{eq:stationary_state}
    x(k)\!=\!\max\!\left(\!0,\lambda\!-\!\mu m_{\alpha,\beta} \left(\!\frac{k}{K}\!\right)^{\!\alpha+1}\!-\!\sigma\sqrt{q_{\alpha,\beta}\left(\!\frac{k}{K}\!\right)^{\alpha+1}}z \right) \,,
\end{equation}
where $z$ is the random variable which follows the standardized normal distribution $\mathcal{N}(0,1)$ with the zero mean and unit variance, and 
\begin{equation}
\label{eq:mq_weighted}
    m_{\alpha,\beta}\equiv 
    \left\langle\left(\frac{k}{K}\right)^{\beta} x \right\rangle_{nn} 
\quad {\rm and} \quad
    q_{\alpha,\beta}=\left\langle\left(\frac{k}{K}\right)^{\beta} x^2 \right\rangle_{nn}
\end{equation}
represent the average ($\langle \cdots \rangle_{nn}$)  of   $\left(\frac{k_j}{K}\right)^{\beta}x_j$ and $\left(\frac{k_j}{K}\right)^{\beta}x_j^2$, respectively, over neighboring species ($j$) and over realizations of interaction strength $\{J_{ij}\}$. 

The interaction strength $J_{ij}$ in Eq.~\eqref{eq:quenchedInteractionStrength} represents the capability of species to interact with other species, which is determined by the intrinsic biological properties of species. Hence, the timescale of interaction strength evolution should be much larger than the timescale of the population dynamics in our model; therefore, we disregard the time evolution of interaction strengths in this model.

\section{Survival probability of species}
\label{sec:survivalProbability}
Let us define a species as viable if its stationary abundance is positive.  The parameters $\alpha$ and $\beta$ control the degree-dependent influence among species. For example, hub species receive a disproportionately stronger incoming impact when $\alpha > 0$, while they exert a stronger outgoing impact on others when $\beta > 0$. Consequently, whether a species is viable depends on its degree through the values of $\alpha$ and $\beta$. To illustrate this effect, we examine the degree-specific survival probability $\phi_{\alpha,\beta}(k)$ via numerical simulations, demonstrating how different parameter combinations $(\alpha, \beta)$ selectively promote the survival of high- or low-degree species. 

For numerical simulations, we construct the Erd\H{o}s--R\'enyi (ER) network~\cite{renyi1959random,gilbert1959random} with the Poisson degree distribution $p(k) = K^{k} e^{-K} / k!$ in the sparse-network limit $K \ll S$, and the static scale-free (SF) networks~\cite{goh2001universal} with the power-law degree distribution $p(k) \sim k^{-\gamma}$ in the tail part $k \gg 1$ where we take two different degree exponents $\gamma = 3.5$ and $\gamma = 2.5$. The number of nodes $S = 4000$ and mean degree $K = 30$ are held fixed throughout this study. Then, we generate the interaction strengths using the transformed random number given by Eq.~\eqref{eq:quenchedInteractionStrength}. Initially, abundances of each species are randomly drawn from the uniform distribution in the range $[0,1]$. Then, we numerically integrate Eq.~\eqref{eq:model} with the parameters $\lambda=0.5$, $\mu=1.0$, and $\sigma=0.3$ by Runge-Kutta-Fehlberg method~\cite{fehlberg1969low} until the stationary state is reached. For each simulation time step, species with a population less than the threshold value $10^{-11}$ were considered extinct, and we set the population of such species to zero. For each parameter set, we use 100 independent realizations for simulations.

After the system has reached the stationary state, we measure the degree-specific survival probability $\phi_{\alpha,\beta}(k)$, which is the fraction of viable species 
with the initial degree $k$ 
for given values of $\alpha$ and $\beta$. Figures~\ref{fig:survivalProbability}(a)--(c) show the survival probability $\phi_{\alpha,\beta}(k)$ for three representative parameter sets $(\alpha,\beta)=(-1.8,-0.4)$, $(-1.0,-0.4)$, and $(0.2,-0.4)$, which exhibit three distinct behaviors in species survival. When $(\alpha,\beta)=(-1.8,-0.4)$, the probability $\phi_{\alpha,\beta}(k)$ is an increasing function of degree $k$, suggesting that species with higher degrees thrive more than low-degree species. On the contrary, for $(\alpha,\beta)=(0.2,-0.4)$, the probability $\phi_{\alpha,\beta}(k)$ decreases with $k$ and species with lower degrees have a greater chance of survival than hub species. Finally, the case with $(\alpha,\beta)=(-1.0,-0.4)$ is a special case where the system shows relatively unbiased survival of species, resulting in $\phi_{\alpha,\beta}(k)$ almost independent of $k$ and all species survive equally. These benchmark cases show that in our model, the survival probability $\phi_{\alpha,\beta}(k)$ can also be biased to promote the survival of hub species as well as low-degree species in the competitive environment, in contrast to the previous study~\cite{park2024incorporating}, corresponding to $(\alpha,\beta)=(0,0)$, which only permits the proliferation of low-degree species in competitive environments with $\mu > 0$.

These biased and unbiased behaviors in $\phi_{\alpha,\beta}(k)$ are also reflected in $p_{\alpha,\beta}^\prime(k)$, the probability distribution of the initial degree that 
viable species originally had (note that the final degree of each node can be decreased by the removal of extinct species at the stationary state), which is calculated as
\begin{equation}
\label{eq:pprime}
    p_{\alpha,\beta}^\prime(k)=\frac{p(k)\phi_{\alpha,\beta}(k)}{\int dk^\prime~p(k^\prime)\phi_{\alpha,\beta}(k^\prime)} \,,
\end{equation}
where $p(k)$ is the degree distribution of the original network.
In Figs.~\ref{fig:survivalProbability}(d)--(f), the $p_{-1,-0.4}^\prime(k)$ (the green lines) show very similar distributions to $p(k)$ of all species (the black lines), not surprisingly from Eq.~\eqref{eq:pprime} with almost constant $\phi_{\alpha,\beta}(k)$ in Figs.~\ref{fig:survivalProbability}(a)--(c). However, $p_{\alpha,\beta}^\prime(k)$ deviates from $p(k)$ in the other cases $(\alpha,\beta)=(-1.8,-0.4)$ and $(0.2,-0.4)$. To be specific, for $(\alpha,\beta)=(-1.8,-0.4)$, the distribution $p_{\alpha,\beta}^\prime(k)$ is above $p(k)$ in the large-$k$ regime and below $p(k)$ when $k$ is small. When $(\alpha,\beta)=(0.2,-0.4)$, conversely, $p_{\alpha,\beta}^\prime(k)$ is slightly above $p(k)$ in the small-$k$ regime, and the tail of $p_{\alpha,\beta}^\prime(k)$ falls much more sharply than $p(k)$ due to the 
low survival probability of hub species. 

This deviation is most prominent in the static SF network with $\gamma = 2.5$‬‬ [Fig.~\ref{fig:survivalProbability}(f)], because its smaller degree exponent yields a more heterogeneous degree distribution with a fatter tail compared to $\gamma = 3.5$‬‬, naturally amplifying these degree-dependent shifts. Note that in the ER random networks [Fig.~\ref{fig:survivalProbability}(d)], these qualitative trends do exist,
but the overall effect is much less prominent. Because the degree variation in ER networks is inherently small, restricting the horizontal axis to a narrow range, these deviations remain visually subdued. 

To identify the different behaviors of $\phi_{\alpha, \beta}(k)$ in a compact fashion, we use the mean initial degree of 
viable species
\begin{equation}
    K_{\alpha,\beta}^\prime=\int dk~kp_{\alpha,\beta}^\prime(k) \,.
\end{equation}
When $\phi_{\alpha, \beta}(k)$ is an increasing function of $k$ (hubs tend to survive), $K_{\alpha,\beta}^\prime>K$. If $\phi(k)$ is a decreasing function of $k$ (hubs tend to go extinct), $K_{\alpha,\beta}^\prime<K$. Finally, $K_{\alpha,\beta}^\prime=K$ (the neutral case) if $\phi_{\alpha, \beta}(k)$ is  constant. In the next section, we expand this analysis to the $\alpha$-$\beta$ space and identify the hub-favored phase where $K_{\alpha,\beta}^\prime>K$, the hub-suppressed phase where $K_{\alpha,\beta}^\prime<K$, and the phase boundary characterized by $K_{\alpha,\beta}^\prime=K$.

\section{The Hub-favored and hub-suppressed phases}
\label{sec:twoPhases}
\begin{figure*}[t]
    \centering
    \includegraphics[width=.9\linewidth]{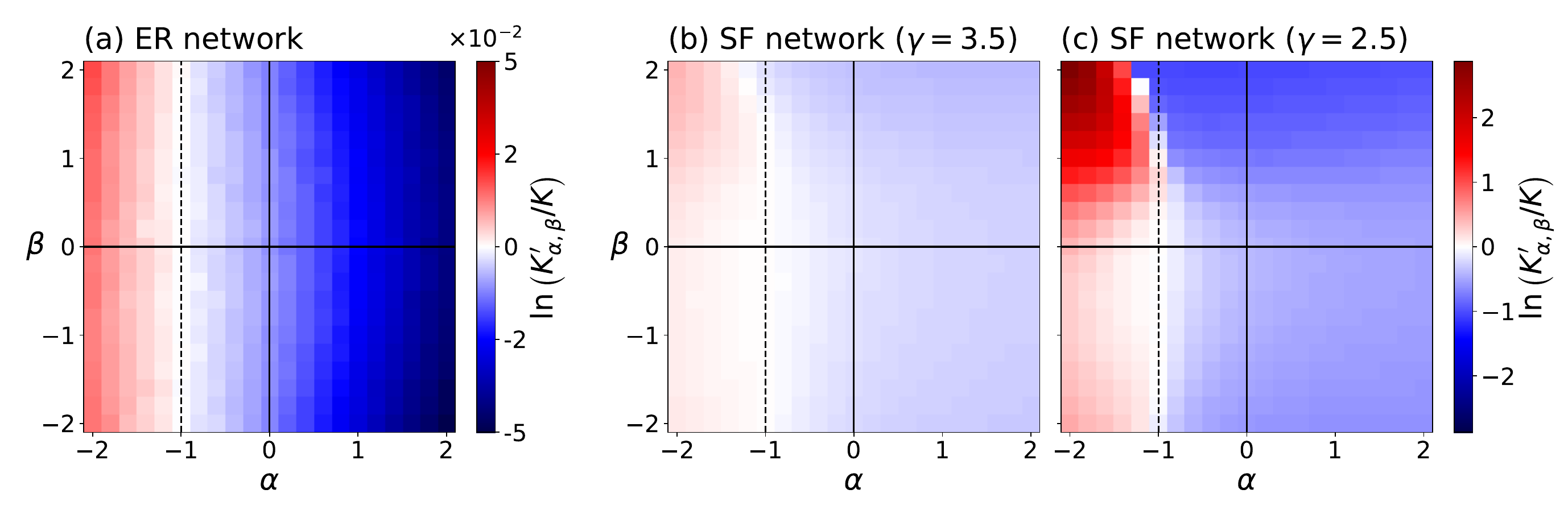}
    \caption{
    The heatmap of $\ln(K_{\alpha,\beta}^\prime/K)$, where $K_{\alpha,\beta}^\prime$ is the mean initial degree of surviving species for a given parameter set $(\alpha,\beta)$, and $K$ is the mean degree of the initial network. Panels (a)--(c) correspond to the ER network, and  static SF networks with $\gamma=3.5$ and $2.5$, respectively. The $(\alpha,\beta)$ plane is divided into a hub-favored phase ($K_{\alpha,\beta}^\prime>K$, red)
    and a hub-suppressed phase ($K_{\alpha,\beta}^\prime<K$, blue).
    The phase boundary is represented by the white region where $K_{\alpha,\beta}^\prime=K$. In the ER network, the phase boundary lies at $\alpha\approx-1$ throughout all values of $\beta$ we consider. Similarly, the phase boundary is also found at $\alpha\approx-1$ in the static SF networks but deviates from $\alpha=-1$ to more negative $\alpha$ values in the large $\beta$ regime. The black dashed vertical line highlights the theoretical prediction ($\alpha=-1$) from the mean-field approach. 
    Each point is averaged over $100$ realizations of the network and interaction strengths.
    }
    \label{fig:survivalDegreeMap_mu_1.0}
\end{figure*}

In the previous section, we have investigated the biased and unbiased behaviors in degree-dependent survival probability $\phi_{\alpha, \beta}(k)$ and suggested that different behaviors of $\phi_{\alpha, \beta}(k)$ are captured by a single measure $K_{\alpha,\beta}^\prime$. Extensive numerical simulations of our model in the competitive environment ($\mu=1.0$) reveal that the model exhibits clearly distinguished hub-favored and hub-suppressed phases in the $\alpha$--$\beta$ space.

Figure~\ref{fig:survivalDegreeMap_mu_1.0} shows the value of $\ln(K_{\alpha,\beta}^\prime/K)$ in the parameter space $(\alpha,\beta)$. For all networks, we observe that the $\alpha$-$\beta$ plane is partitioned into the hub-favored phase (the red region) where $K_{\alpha,\beta}^\prime>K$ and the hub-suppressed phase (the blue region) where $K_{\alpha,\beta}^\prime<K$. In the hub-favored phase, hub species with large degrees show larger survival probability than species with smaller degrees. On the other hand, exactly opposite behavior is observed in the hub-suppressed phase. 
In that phase, lower-degree species have a higher chance of survival than species with larger degrees.
The boundary between these phases is marked by the white region, which corresponds to the region where $K_{\alpha,\beta}^\prime=K$ is satisfied.

In the ER network, the phase boundary lies at the vertical region $\alpha\approx -1$ as shown in Fig.~\ref{fig:survivalDegreeMap_mu_1.0}(a). Similarly, the phase boundary is also at $\alpha\approx -1$ in the static SF network, but the phase boundary deviates from $\alpha\approx -1$ and moves towards more negative $\alpha$ in the large $\beta$ regime [Figs.~\ref{fig:survivalDegreeMap_mu_1.0}(b)--(c)]. 
The phase boundary 
can be understood from the total incoming influence  
$I^{\leftarrow}(k)$ of a species with degree $k$, defined in Eq.~\eqref{eq:incoming}. 
Under the mean-field approach in Appendix~\ref{app:mean-field}, its expectation value is given  by 
\begin{equation}
\label{eq:totalIncomingStrength}
    \langle I^{\leftarrow}(k)\rangle\approx\mu \left(\frac{k}{K}\right)^{\alpha+1} m_{\alpha,\beta},
\end{equation}
where $m_{\alpha,\beta}$ characterizes the properties of a neighbor species, as  defined in Eq.~\eqref{eq:mq_weighted}. 
Suppose that $m_{\alpha,\beta}$ and $q_{\alpha,\beta}$ are independent of the degree $k$ of the focal species. Then 
$\langle I^\leftarrow(k) \rangle$ becomes independent of $k$ at $\alpha=-1$. Since the expectation value of the stationary abundance satisfies 
\begin{equation}
    \langle x(k)\rangle = \max(0,\lambda - \langle I^{\leftarrow}(k)\rangle)
    \label{eq:xk_avg}
\end{equation} 
according to Eq.~\eqref{eq:stationary_state},  $\langle x(k)\rangle$ is also independent of $k$ at $\alpha=-1$. 

The phase boundary at $\alpha=-1$ can also be understood through the derivative of $\phi_{\alpha, \beta}(k)$ with respect to $k$, because it determines whether $\phi_{\alpha, \beta}(k)$ is increasing or decreasing and thus the sign of $\ln(K_{\alpha,\beta}^\prime/K)$, as discussed in Sec.~\ref{sec:survivalProbability}. 
The stationary abundance of species with degree $k$ follows the conditional distribution $\rho(x|k)$ derived from Eq.~\eqref{eq:stationary_state}. This distribution exhibits a truncated Gaussian distribution whose peak (technically different from the mean as it is truncated) is at $\langle x(k)\rangle$,  given in Eq.~\eqref{eq:xk_avg}, 
while its variance (before truncation) is $\sigma^2 q_{\alpha,\beta}(k/K)^{\alpha+1}$ [see Eq.~\eqref{eq:rho_x_k} for the explicit definition].
The survival probability $\phi_{\alpha, \beta}(k)$ is then written by
\begin{align}
    \phi_{\alpha, \beta}(k)&=\int_{0^{+}}^\infty dx~\rho(x|k) =\frac{1}{\sqrt{2\pi}}\int_{-\infty}^{\Delta(k)} dz ~e^{-z^2/2}
\end{align}
with 
\begin{equation}
\label{eq:delta}
    \Delta(k)=\frac{\lambda-\mu m_{\alpha,\beta} (k/K)^{\alpha+1}}{\sigma\sqrt{q_{\alpha,\beta} (k/K)^{\alpha+1}}}. 
\end{equation}
The $k$ dependence of $\phi_{\alpha,\beta}(k)$ can be understood by its  derivative  with respect to $k$ given by 
\begin{align}
    \frac{\partial \phi_{\alpha, \beta}(k)}{\partial k}=\frac{1}{\sqrt{2\pi}}\frac{\partial\Delta(k)}{\partial k}e^{-\Delta^2\!(k)/2}.
\end{align}
The hubs are favored to survive if $\partial\phi_{\alpha, \beta}(k)/\partial k>0$ (hub-favored phase). On the other hand, if $\partial\phi_{\alpha, \beta}(k)/\partial k<0$, the hub species have higher probabilities to go extinct (hub-suppressed phase). The phase boundary is the set of solutions satisfying $\partial\phi_{\alpha, \beta}(k)/\partial k=0$.  
As $e^{-\Delta(k)^2/2}$ is always positive, 
the phase boundary is determined by $\partial\Delta(k)/\partial k$, which is evaluated as
\begin{equation}
\label{eq:deltadiff}
    \frac{\partial \Delta(k)}{\partial k}=-\frac{\alpha+1}{2 \sigma K \sqrt{q_{\alpha,\beta}}}\left[\lambda\!\left(\frac{k}{K}\right)^{\!-\frac{\alpha+3}{2}}\!+\!\mu m_{\alpha,\beta}\left(\frac{k}{K}\right)^{\!\frac{\alpha-1}{2}}\right].
\end{equation}
The sign of $\partial\Delta(k)/\partial k$ is identical to the sign of $-(\alpha+1)$ because other parts are positive in the competitive environment ($\mu>0$). Therefore, we can verify that the phase boundary is at $\alpha=-1$: when $\alpha=-1$, $\partial\Delta(k)/\partial k=0$ holds for all $k$, so the survival probability $\phi_{\alpha,\beta}(k)$ becomes independent of degree $k$, marking the boundary at which the sign of $\partial\Delta(k)/\partial k$ is flipped.
When $\alpha>-1$, $\partial\phi_{\alpha, \beta}(k)/\partial k<0$, and the survival probability $\phi_{\alpha,\beta}(k)$ monotonically decreases with the degree $k$; the system is in the hub-suppressed phase. On the other hand, the system is in the hub-favored phase when $\alpha<-1$.

For SF networks, more notably when $\gamma = 2.5$, 
where the severe degree heterogeneity leads the second moment $\langle k^2 \rangle = \int dk \, k^2 p(k)$ diverge in the thermodynamic limit,
the phase boundary deviates from $\alpha=-1$ and bends toward more negative values of $\alpha$ as $\beta$ grows [Figs.~\ref{fig:survivalDegreeMap_mu_1.0}(b) and (c)]. 
This deviation can be understood qualitatively  from the growing heterogeneity of neighbor contributions when $\beta$ is large. 

The phase boundary at $\alpha=-1$ is based on the assumption that $m_{\alpha,\beta}$ and $q_{\alpha,\beta}$ are independent of the degree $k$ of a focal node, which can be violated in SF networks for large $\beta$. Neglecting degree-degree correlation, $m_{\alpha, \beta}$ is evaluated in terms of the edge-based degree distribution $p_{nn}(k')\sim k'^{1-\gamma}$ as 
\begin{equation}
\label{eq:m_alpha_beta_calc}
    m_{\alpha,\beta}
    \simeq \langle x \rangle\int^{k_\text{max}(k)} dk'~p_{nn}(k')\left(\frac{k'}{K}\right)^{\beta} 
    \sim \int^{k_\text{max}(k)} dk' k'^{1+\beta-\gamma},
\end{equation}
and $q_{\alpha,\beta}$ are evaluated similarly, where $k_\text{max}(k)$ is the largest degree among the $k$ neighbors of a degree-$k$ node.  Also we used that, at the phase boundary,  $\langle x(k)\rangle$ is independent of $k$ and equal to the global average $\langle x\rangle$. While the integral is dominated by its lower end for $\beta<\gamma-2$, it grows as $k_\text{max}(k)^{2+\beta-\gamma}$ for $\beta>\gamma-2$. Since $k_\text{max}(k)$ satisfies the extreme-value condition that one neighbor is expected above it, i.e., $\int_{k_\text{max}(k)} dk' p_{nn}(k') \simeq 1/k$, it scales as $k_\text{max}(k)\sim k^{1/(\gamma-2)}$. Consequently, $m_{\alpha,\beta}$ and $q_{\alpha,\beta}$ scale as $\sim k^{(2+\beta-\gamma)/(\gamma-2)}$ for $\beta>\gamma-2$ while they are $k$-independent for $\beta<\gamma-2$. Hence, for $\beta>\gamma-2$,  the total incoming influence in Eq.~\eqref{eq:totalIncomingStrength} is dominated by the largest-degree neighbor, $\langle I^\leftarrow (k)\rangle /(k/K)^\alpha \propto k_\text{max}(k)^\beta$, pushing the phase boundary  to more negative values, whereas for $\beta<\gamma-2$ it is contributed by all neighbors, yielding $\langle I^\leftarrow(k)\rangle/ (k/K)^\alpha \propto k$  and leaving the boundary at $\alpha=-1$.

This estimate, however, overestimates the shift. It predicts, for instance, the boundary at $\alpha=-\beta/(\gamma-2)$ for large $\beta$, whereas the simulations suggest a smaller deviation from $\alpha=-1$.  This is largely due to a finite-size effect. A neighbor's degree cannot exceed the global maximum degree scaling as $S^{1/(\gamma-1)}$. Consequently, $k_\text{max}(k)$ rapidly saturates at the finite-size cutoff, substantially weakening the $k$-dependence of $m_{\alpha,\beta}$ and $q_{\alpha,\beta}$ and thereby pulling the phase boundary back toward $\alpha=-1$. Nevertheless, the argument captures why the deviation from $\alpha=-1$ grows with $\beta$ and with decreasing $\gamma$. A quantitative theory of this shift remains an interesting direction for future work.

\section{Evolution of degree assortativity in the hub-favored and hub-suppressed phases}
\label{sec:assortativity}
\begin{figure}[t]
    \centering
    \includegraphics[width=\linewidth]{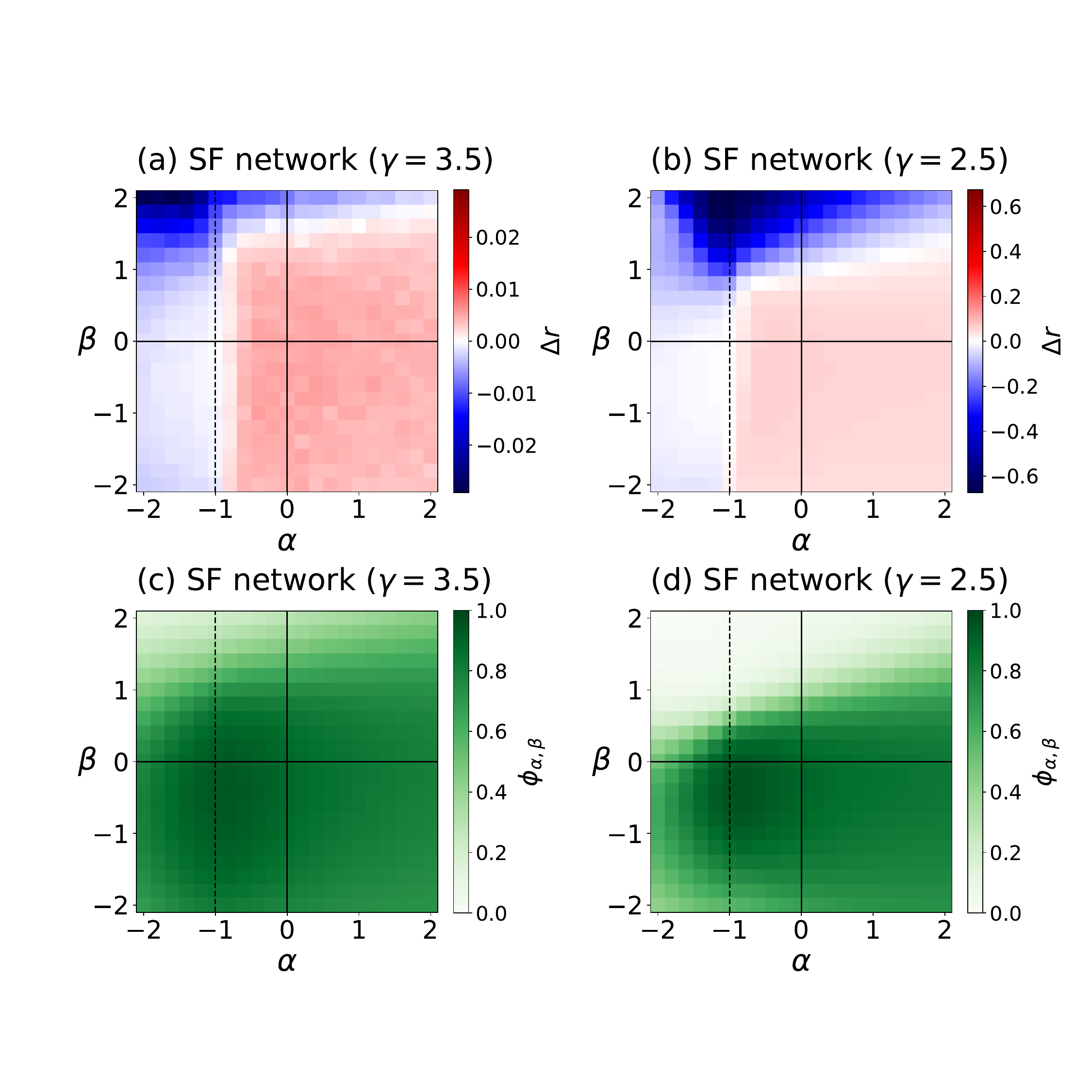}
    \caption{(a), (b) Change in the degree assortativity coefficient $\Delta r$ and (c), (d) mean survival probability $\langle \phi_{\alpha,\beta} \rangle$ in the $\alpha$-$\beta$ plane for static SF networks with (a), (c) $\gamma = 3.5$ and (b), (d) $\gamma = 2.5$. The black dashed vertical line indicates the phase boundary at $\alpha = -1$ predicted by the mean-field approach. The degree assortativity of the surviving network decreases in the hub-favored phase ($\alpha < -1$) while it increases in the hub-suppressed phase ($\alpha > -1$), except in the large-$\beta$ regime where the mean survival probability becomes very low.
    Each point is averaged over 100 simulations with the same simulation settings as in Sec.~\ref{sec:twoPhases}.}
    \label{fig:degreeAssortativityChangeMap_mu_1.0}
\end{figure}

In the hub-favored or hub-suppressed phases, the survival of species depends on $k$, altering the degree distribution of surviving species and thereby shifting network-level properties, including not only the mean degree $K$ but also degree assortativity $r$, which is the Pearson correlation coefficient between the excess degrees of nodes at either end of an edge~\cite{newman2003mixing}. To investigate this shift induced by $k$-dependent survival, we examine how degree assortativity evolves from its initial value. 

The change in the degree assortativity coefficient is denoted by $\Delta r = r_f - r_i$, where $r_i$ is the degree assortativity coefficient of the initial network and $r_f$ is that of the network only composed of surviving  species formed after the system has reached its stationary state. When $\Delta r > 0$, the surviving network becomes more assortative, indicating that edges connecting hub and non-hub species 
are preferentially removed as a result of extinct species. 
Conversely, $\Delta r < 0$ indicates a shift toward disassortativity, where the edge removal process occurs between nodes with similar degree values. 

For static SF networks with $\gamma = 3.5$ and $2.5$, the resulting $\Delta r$ is shown in Figs.~\ref{fig:degreeAssortativityChangeMap_mu_1.0}(a) and (b). Different behaviors in the degree-specific survival probability $\phi_{\alpha,\beta}(k)$ in the hub-favored and hub-suppressed phases lead to opposite trends in the degree assortativity coefficient. In the hub-suppressed phase ($\alpha > -1$), hub nodes are preferentially removed, eliminating most hub-to-non-hub interactions that had originally contributed to the degree disassortativity, so $\Delta r > 0$. Conversely, in the hub-favored phase, low-degree species are preferentially removed, preserving most hub-to-non-hub interactions, so $\Delta r < 0$.

This tendency, however, does not hold in the large-$\beta$ regime, where degree assortativity exhibits a sharp decline. In this regime, the mean survival probability $\langle \phi_{\alpha,\beta}\rangle=\int dk~p(k)\phi_{\alpha,\beta}(k)$ is too low [see Figs.~\ref{fig:degreeAssortativityChangeMap_mu_1.0}(c) and (d)]: only a few hundred species survive out of $S = 4000$. Such large-scale extinction 
leaves only a small subset of species and interactions, making the assortativity highly sensitive to the detailed composition of the remaining edges. Consequently, the simple interpretation based on the preferential removal of hubs or non-hubs is no longer sufficient.
Nevertheless, in the moderate-$\beta$ regime where the survival probability remains sufficiently high, the observed trends in degree assortativity are in good agreement with our interpretation.

\section{Summary and Discussion}
\label{sec:summary}
In this study, we have introduced degree-dependent interaction strength into the generalized random Lotka-Volterra (GRLV) model and investigated its effect on the survival probability of species with different degrees. We have found that the correlation between interaction strength and degree determines the survivability across species, giving rise to two distinct phases: the hub-favored phase, in which the degree-specific survival probability $\phi(k)$ increases with $k$, and the hub-suppressed phase, in which $\phi(k)$ decreases with $k$. Furthermore, we showed via mean-field (MF) approximation that the phase boundary is located at $\alpha = -1$, consistent with simulation results except at large values of $\beta$. At $\alpha = -1$, the degree dependence of interaction strength exactly compensates for the difference in the number of neighbors, so all species experience the same net incoming effect regardless of degree. At large values of $\beta$, however,  the growing heterogeneity of neighbor contributions causes a deviation from 
the MF approximation in SF networks. 

Beyond the survival statistics, the two phases also leave a distinct imprint on the network structure of surviving species. In the hub-favored phase, hub species survive preferentially while low-degree species are eliminated. The interactions between hub and non-hub species are therefore largely preserved, whereas interactions between low-degree species are selectively removed. Since the latter contribute to degree assortativity, their loss causes $r$ to decrease, enhancing degree disassortativity. In the hub-suppressed phase, the opposite occurs: hub species are disproportionately removed, together with the hub-to-non-hub interactions that are the primary source of degree disassortativity. Consequently, $r$ increases and the network becomes less disassortative. 
Nevertheless, the increase is not sufficient to reverse the intrinsic disassortative structure of the underlying scale-free network, so the final network remains 
disassortative ($r_f < 0$) in both phases.

To connect these findings to real ecological systems, we measured the degree assortativity coefficient $r$ in empirical food web data~\cite{jaarsma1998characterising,thompson1999effect,townsend1998disturbance,thompson2000resolution,thompson2001allocation,thompson2003impacts,thompson2005energy}, which encodes predator-prey interactions among species. These networks exhibit substantial degree disassortativity: the mean assortativity coefficient is $\langle r \rangle \approx -0.295$ with a standard deviation of $\sqrt{\langle r^2 \rangle - \langle r \rangle^2} = 0.093$. Given that static scale-free networks already display negative degree assortativity for $\gamma = 2.5$~\cite{lee2006intrinsic}, our results indicate that the hub-favored phase further amplifies this disassortativity, whereas the hub-suppressed phase partially reduces it. The close correspondence between the hub-favored scenario and the empirical food web data supports the hypothesis that hub species tend to have weaker interaction strengths, as suggested in~\cite{koch2026many,gellner2023stable}, an effect that reinforces degree disassortativity in ecological networks.

One direction for future work is a node-removal approach. The hub-favored and hub-suppressed phases correspond to hub-protecting and hub-targeted extinction scenarios, respectively, since species extinction can be interpreted as node removal in ecological networks. It is well established that node removal strategies alter key network properties, including the degree distribution~\cite{lee2022degree}, the size of the largest connected cluster~\cite{cohen2001breakdown,cohen2000resilience}, and the degree assortativity coefficient~\cite{wang2014emergence}. Thus, from the perspective of node removal, studying secondary extinctions~\cite{allesina2006secondary,ebenman2004community}---quantifying how the population of species with degree $k$ responds when a species of degree $k'$ is removed---would help identify keystone species and their systemic role. Such an approach would bridge the gap between theoretical models of species coexistence and the structural properties observed in empirical ecological networks.

\begin{acknowledgments}
This research was supported by the National Research Foundation of Korea (NRF) grant Nos. RS-2026-25497103, RS-2024-00460958 (H.J.P.) and RS-2026-25468383 (S.H.L.) funded by the Korea government (MSIT), and by KIAS Individual Grant No. CG079902 (D.-S.L.) at Korea Institute for Advanced Study. S.S. acknowledges Iniziativa PNC0000002- DARE – Digital Lifelong Prevention and INFN Lincoln grant.
\end{acknowledgments}

\appendix

\section{Derivation of Eq.~\eqref{eq:stationary_state}}
\label{app:mean-field}
We obtain the stationary solution of 
Eq.~\eqref{eq:model} using the mean-field approach~\cite{galla2018dynamically,bunin2017ecological,roy2019numerical,park2024incorporating}. 
The total incoming influence in Eq.~\eqref{eq:model} can be divided into two terms
\begin{equation}
\label{eq:interaction}
    \sum_{j\neq i}J_{ij}A_{ij}x_j(t)=\sum_{j\neq i}\mu_{ij} A_{ij}x_j(t) 
    +\sum_{j\neq i}\sigma_{ij}z_{ij} A_{ij}x_j(t) \,,
\end{equation}
where the first term on the right-hand side accounts for the mean interaction strengths in Eq.~\eqref{eq:interaction_mean} and the second term describes the fluctuation as given in Eq.~\eqref{eq:interaction_variance}. 
The first term can be represented by 
\begin{align}
\label{eq:firstterm}
    \sum_{j\in nn(i)}\mu_{ij} x_j(t)
    &=\mu \left({k_i \over K}\right)^{\alpha+1}{1\over k_i}\sum_{j\in nn(i)}  \left(\frac{k_j}{K}\right)^{\beta} x_j(t) \nonumber\\
    &=\mu\left({k_i \over K}\right)^{\alpha+1} \left\langle \left(\frac{k}{K}\right)^{\beta}x(t) \right\rangle_{nn(i)}
    \, ,
\end{align}
where $nn(i)$ denotes the set of interacting neighbors of species $i$ and $\langle f\rangle_{nn(i)} = k_i^{-1}\sum_{j\in nn(i)}f_j$ is the average over the neighbors of $i$. 
The second term in Eq.~\eqref{eq:interaction} can be represented as 
\begin{align}
\label{eq:influence_fluct}
    \sum_{j\in nn(i)}\sigma_{ij} z_{ij} x_j(t) 
    &=\sqrt{\frac{\sigma^2}{K}}\sqrt{\left(\frac{k_i}{K}\right)^\alpha}\xi_i(t) \,,
\end{align}
where $\xi_i(t)$ is a random variable defined by
\begin{equation}
    \xi_i(t)\equiv \sum_{j\in nn(i)} \sqrt{ \left(\frac{k_j}{K}\right)^\beta} z_{ij} x_j(t) \,,
\end{equation}
and $z_{ij}$ is the standardized Gaussian random variable with $\langle z_{ij} \rangle=0$, and $\langle z_{ij} z_{i^\prime j^\prime}\rangle=\delta_{ii^\prime}\delta_{jj^\prime}$. Here $\langle \cdots\rangle$ denotes the average over different realizations of $\{J_{ij}\}$.

The first and second moments of $\xi_i$ can be obtained as follows. 
The mean of $\xi_i(t)$ is
\begin{align}
\label{eq:xi_mean}
    \langle\xi_i(t)\rangle=\sum_{j\neq i} \sqrt{ \left(\frac{k_j}{K}\right)^\beta} \langle z_{ij}\rangle A_{ij}x_j(t) 
    =0 \,.
\end{align}
Likewise, the second moment of $\xi_i(t)$ is
\begin{align}
\label{eq:xi_covariance}
    \langle\xi_i(t)\xi_{i^\prime}(t)\rangle&=\left\langle \sum_{j\in nn(i)} \sqrt{ \left(\frac{k_j}{K}\right)^\beta} z_{ij}x_j(t) \right. \nonumber 
    \left.
     \sum_{j^\prime \in nn(i^\prime)}\sqrt{ \left(\frac{k_{j^\prime}}{K}\right)^\beta} z_{i^\prime j^\prime} x_{j^\prime}(t) \right\rangle  \nonumber \\
    &=\delta_{ii^\prime}\sum_{j\in nn(i)} \left\langle \left(\frac{k_j}{K} \right)^\beta x_j^2(t) \right\rangle
    =\delta_{ii^\prime}k_i\left\langle \left(\frac{k}{K}\right)^{\beta}x^2(t)\right\rangle_{nn(i)} \, .
\end{align}

Utilizing Eqs.~\eqref{eq:firstterm}, ~\eqref{eq:influence_fluct}, ~\eqref{eq:xi_mean}, and ~\eqref{eq:xi_covariance}, we get the stationary abundance of species with degree $k$ as
\begin{equation}
\label{eq:app:stationary_state}
x(k)=\max\left(0,\lambda-\mu m_{\alpha,\beta} \left(\frac{k}{K}\right)^{\alpha+1}-\sigma\sqrt{q_{\alpha,\beta}\left(\frac{k}{K}\right)^{\alpha+1}}z \right) \,,
\end{equation}
where $z$ is the random number drawn from the standardized distribution,   and
\begin{equation}
\label{eq:m_alpha_beta}
m_{\alpha,\beta}=\left\langle\left(\frac{k}{K}\right)^{\beta} x \right\rangle_{nn} \,,
\end{equation}
and
\begin{equation}
\label{eq:q_alpha_beta}
q_{\alpha,\beta}=\left\langle\left(\frac{k}{K}\right)^{\beta} x^2 \right\rangle_{nn} \, 
\end{equation}
with $\langle \cdots \rangle_{nn}$ denoting the average over neighboring species and over realizations of interaction strengths. 

Since $z$ in Eq.~\eqref{eq:stationary_state} or equivalently Eq.~\eqref{eq:app:stationary_state} follows the standardized distribution $p(z) = (2\pi)^{-1/2} e^{-z^2/2}$, the stationary abundance of species with degree $k$ follows the Gaussian distribution whose mean is $x_0=\lambda-\mu m_{\alpha,\beta}(k/K)^{\alpha+1}$ and standard deviation is $\sigma_0=\sigma\sqrt{q_{\alpha,\beta}(k/K)^{\alpha+1}}$ if the abundance is positive.  Therefore the stationary  abundance distribution of species with degree $k$ can be written as
\begin{equation}
    \rho(x|k)=
    \begin{cases}
    \frac{1}{\sqrt{2\pi}\sigma_0}e^{-\frac{(x-x_0)^2}{2\sigma_0^2}} & (x>0), \\
    A & (x=0),\\
    0 & (x<0),
    \end{cases}
    \label{eq:rho_x_k}
\end{equation}
where $A=\int_{-\infty}^{0} dx \frac{1}{\sqrt{2\pi}\sigma_0} e^{-\frac{(x-x_0)^2}{2\sigma_0^2}}$ is the normalization factor that guarantees $\int_{-\infty}^\infty dx \rho(x|k)= 
1$, satisfying $A=1-\phi_{\alpha, \beta}(k)$. We note that Eq.~\eqref{eq:app:stationary_state}, Eq.~\eqref{eq:m_alpha_beta}, and Eq.~\eqref{eq:q_alpha_beta} restore the stationary state solution in Ref.~\cite{park2024incorporating} when $\alpha=\beta=0$.

\bibliography{bibliography}

\end{document}